\documentclass{article}
\pdfoutput=1
\usepackage{graphicx} 
\graphicspath{ {./figures/} }
\usepackage{arxiv}

\usepackage[utf8]{inputenc} 
\usepackage[T1]{fontenc}    
\usepackage{hyperref}       
\usepackage{url}            
\usepackage{booktabs}       
\usepackage{amsfonts}       
\usepackage{nicefrac}       
\usepackage{microtype}      
\usepackage{lipsum}
\usepackage{lineno}
\usepackage{caption}
\usepackage{subcaption}
\usepackage{amsmath}
\usepackage{authblk}
\usepackage{xcolor}
\usepackage{verbatim}   
\newcommand\correspondingauthor{\thanks{sourav.tarafdar@vanderbilt.edu}}
\title{Effective Gain and Ion Back Flow study of triple and quadruple GEM detector}
\author[]{Senta V. Greene}
\author[]{Julia Velkovska}
\author[]{Brandon Blankenship}
\author[]{Michael Z. Reynolds}
\author[]{Sourav Tarafdar\correspondingauthor{}}
\affil[]{(Vanderbilt University, Nashville, U.S.A)}

\begin{document}
\maketitle


\begin{abstract}
Gas electron Multipliers (GEM) are a new generation of
gaseous avalanche devices in the Micro Patterned Gaseous Detector
category. GEMs are widely used in both nuclear and high energy experiments as well as in medical
science. Several parameters define the performance of these types of devices
under various experimental conditions. This article focuses on the study of effective gain and Ion Back Flow (IBF) in both triple and quadruple GEM detectors. Effective gain and IBF are two of the most important parameters in determining the performance of GEM detectors. 

\end{abstract}

\keywords{GEM \and Effective Gain \and IBF \and MPGD}

\section{Introduction}
In recent years, Gas Electron Multipliers (GEMs)~\cite{GEM_sauli} have been used extensively in several high
energy and heavy ion experiments \textcolor{black} {due to their excellent rate capability and intrinsic position resolution. Some of these examples are} the Hadron Blind Detector~\cite{AIDALA2003200} in the PHENIX detector at the Relativistic Heavy Ion Collider (RHIC) at Brookhaven National Laboratory, the ALICE detector at the Large Hadron Collider at \textcolor{black}{CERN}, and the Time Projection Chamber (TPC) for tracking charged particles in the sPHENIX detector under construction at RHIC~\cite{GARABATOS2004197,Tarafdar:2019oB}. GEMs are found to be radiation hard, capable of amplifying small signals, have excellent energy resolution, and deliver fast signal ~\cite{GEM_sauli2}. Some of the parameters that
determine the performance of GEM detectors are the effective gain, energy resolution, spatial resolution, time resolution and ion back flow (IBF). Important factors that determine the effective functioning of GEM detectors are the operating voltages, gas medium in which
the detector is being operated, readout board segmentation, and pitch of the holes on the GEM layers. This
article focuses on studying effective gain and IBF using different gas mixtures and different voltage configurations for both triple and quadruple GEM detectors. Further, two methods for
estimating the effective gain of GEM detectors are also discussed. 
\section{Experimental Configuration}
\begin{figure}[hbt!]
     \centering
     \begin{subfigure}[hbt!]{0.3\textwidth}
         \centering
         \includegraphics[width=\textwidth]{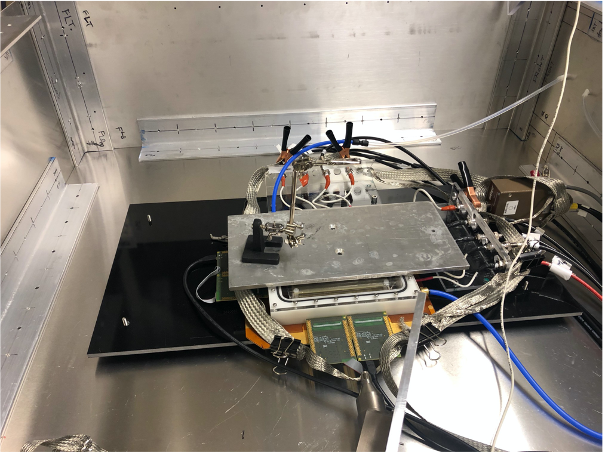}
         \caption{}
         \label{fig:picture}
     \end{subfigure}
     \hfill
     \begin{subfigure}[hbt!]{0.3\textwidth}
         \centering
         \includegraphics[width=\textwidth]{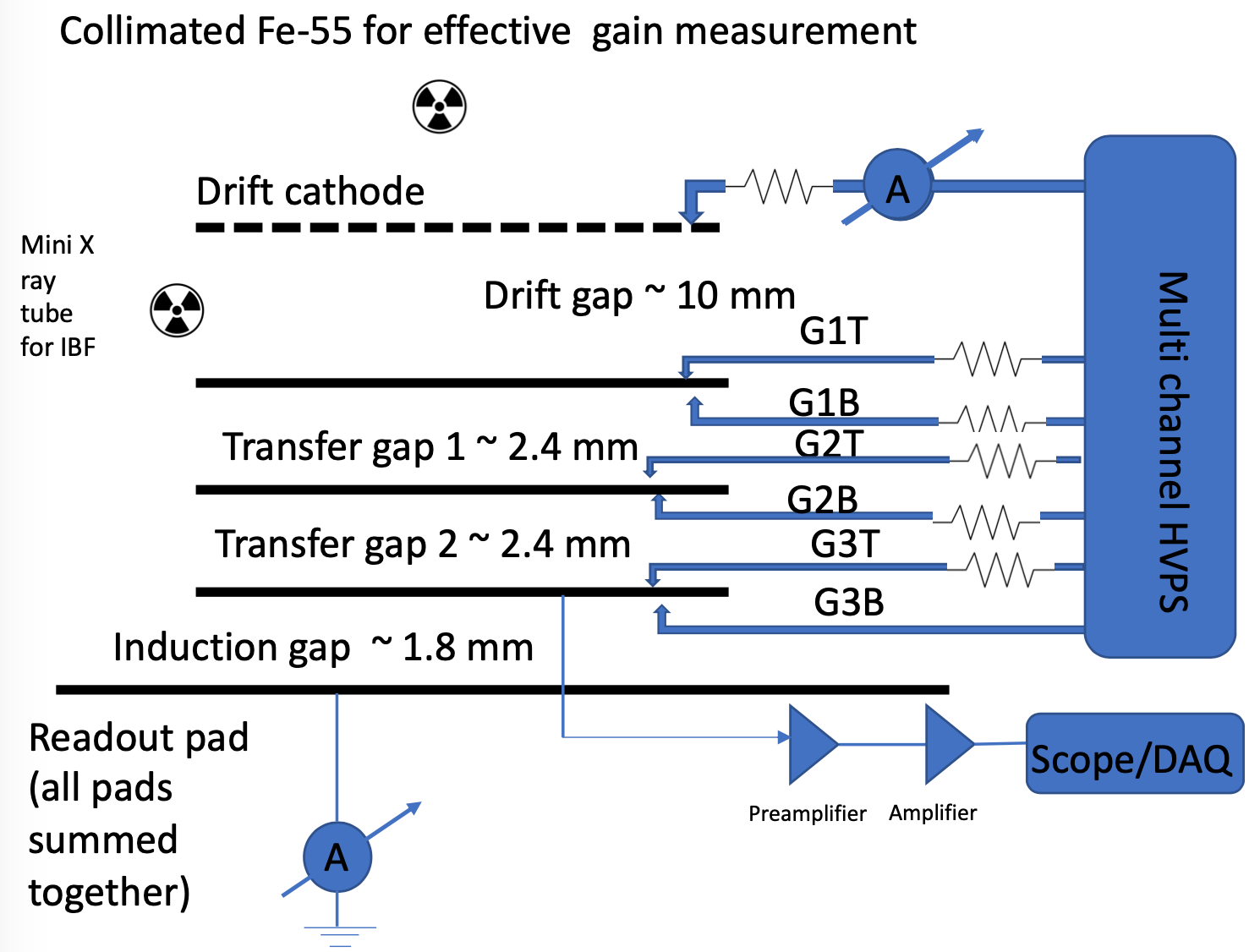}
         \caption{}
         \label{fig:triple_GEM}
     \end{subfigure}
     \hfill
     \begin{subfigure}[hbt!]{0.3\textwidth}
         \centering
         \includegraphics[width=\textwidth]{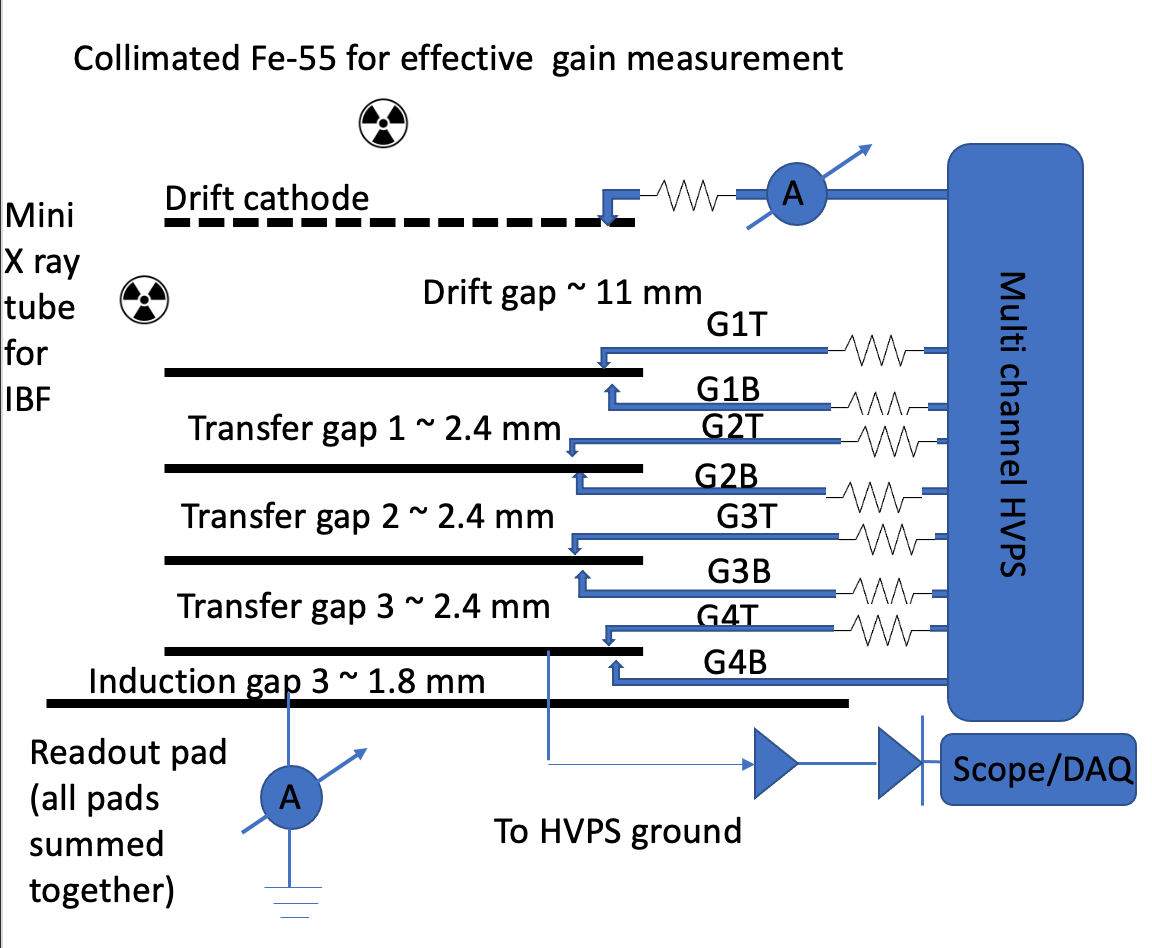}
         \caption{}
         \label{fig:quad_GEM}
     \end{subfigure}
        \caption{Experimental set up a) Picture of GEM detector on experimental test bench b) Schematics of triple GEM detector set up c) Schematics of quadruple GEM detector set up}
        \label{fig:setup}
\end{figure}

The GEM foils used in this study are 
\textcolor{black}{double mask standard GEMs from CERN} with 140 micron pitch and
70 micron hole diameter with $10\times 10~{\rm cm}^{2}$ active area\textcolor{black}{~\cite{GDD}}. The triple GEM detector is composed of a drift cathode of $10\times10~{\rm cm}^{2}$ active area,
three layers of GEMs, and a 512 channel X-Y
strip readout board. The quadruple GEM 
detector has four layers of stacked GEMs \textcolor{black}{in between} the drift 
cathode and 512 channel X-Y strip readout board. \textcolor{black}{The gaps between GEM leyers is known as transfer gap while between GEM foil and readout board is induction gap. For triple GEM detector there are two tansfer gaps while for quadruple GEM detector there are three transfer gaps. Both triple and quadruple GEM detectors has one induction gap.} Fig.~\ref{fig:picture} shows the picture of the experimental
configuration on the laboratory test bench and Figures ~\ref{fig:triple_GEM} and ~\ref{fig:quad_GEM}  show
the schematics of the overall set up for triple and
quadruple GEM detectors, respectively, for measuring effective gain and IBF. The measurement of effective gain for both triple
and quadruple GEM was done in Ar:CO$_{2}$(70:30) and Ar:CO$_{2}$(80:20)
gas mixtures while the optimization of IBF study was performed only using the
Ar:CO$_{2}$(70:30) gas mixture. The gas mixtures used in the
experiment were mixed with an in-house-built gas mixing system,
capable of mixing two different gases with good precision. Both Ar and CO$_{2}$ gas were of ultra high purity (99.999$\%$ pure). \textcolor{black}{The effective gain scanning for both triple and quadruple GEM detectors were done by biasing the all the electrodes of detector using single channel of  high voltage power supply via voltage divider. For IBF optimization 
both} the triple and quad GEM detectors individual electrodes were biased
using individual channels of a high voltage power supply via 10 M$\Omega$ protection
resistors. For measuring the effective gain of the detector, a collimated
monochromatic $^{55}{\rm Fe}$ radioactive source was used, whereas for measuring IBF, a mini
X-ray tube with 1 mm collimation was used. The whole  set up was placed in an
Aluminum box for shielding from X-rays during operation. The
readout electronics for the detector signal included a charge-sensitive \textcolor{black} {Ortec preamplifier with rise time of $<$ 525 ns ~\cite{ORTEC_PA}}, Ortec shaping amplifier and oscilloscope with the capability of
producing histograms of the detected signals. For estimating IBF, the induced current in the readout board from avalanche electrons and that from the drift
cathode due to back-drifting positive ions were measured using a highly-sensitive \textcolor{black}{Picologic} picoammeter~\cite{UTROBICIC201521}. \textcolor{black}{Digital humidity and temperature controllers were used for keeping the atmospheric temperature at 26$^{\circ}$C and relative humidity at 30$\%$.}

\section{Experimental Results}
\label{sec:results}
\subsection{Effective Gain}
\label{subsec:effgain}
\begin{figure}
     \centering
     
     \begin{subfigure}[h!]{0.5\textwidth}
         \centering
       \includegraphics[width=9 cm, height = 4.0 cm]{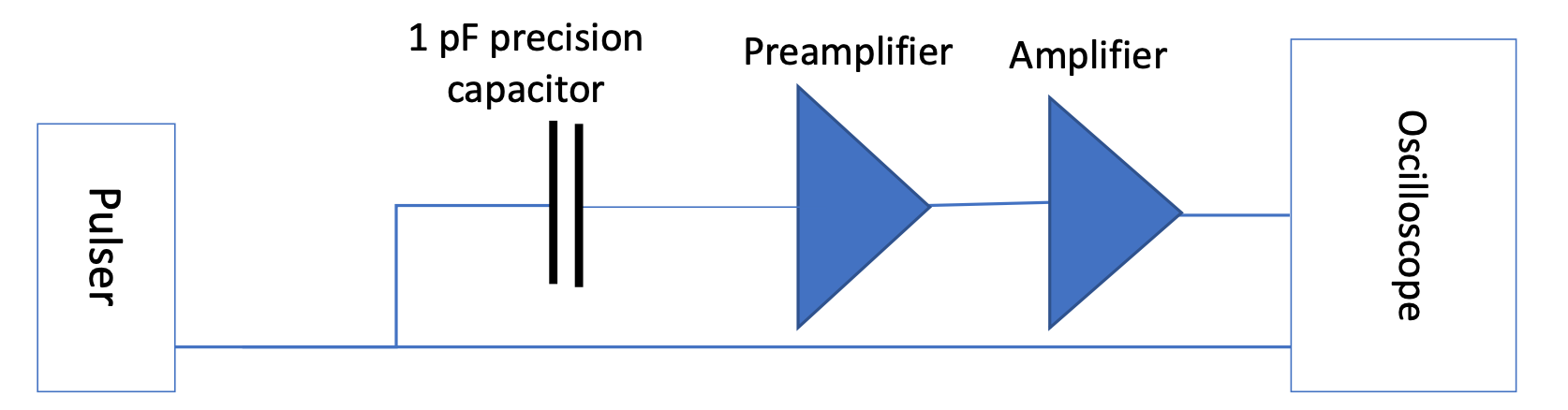}
         \caption{}
         \label{fig:cal_scheme}
     \end{subfigure}
     \hfill
     \begin{subfigure}[h]{0.4\textwidth}
         \centering
         \includegraphics[width=6 cm, height = 3.9 cm]{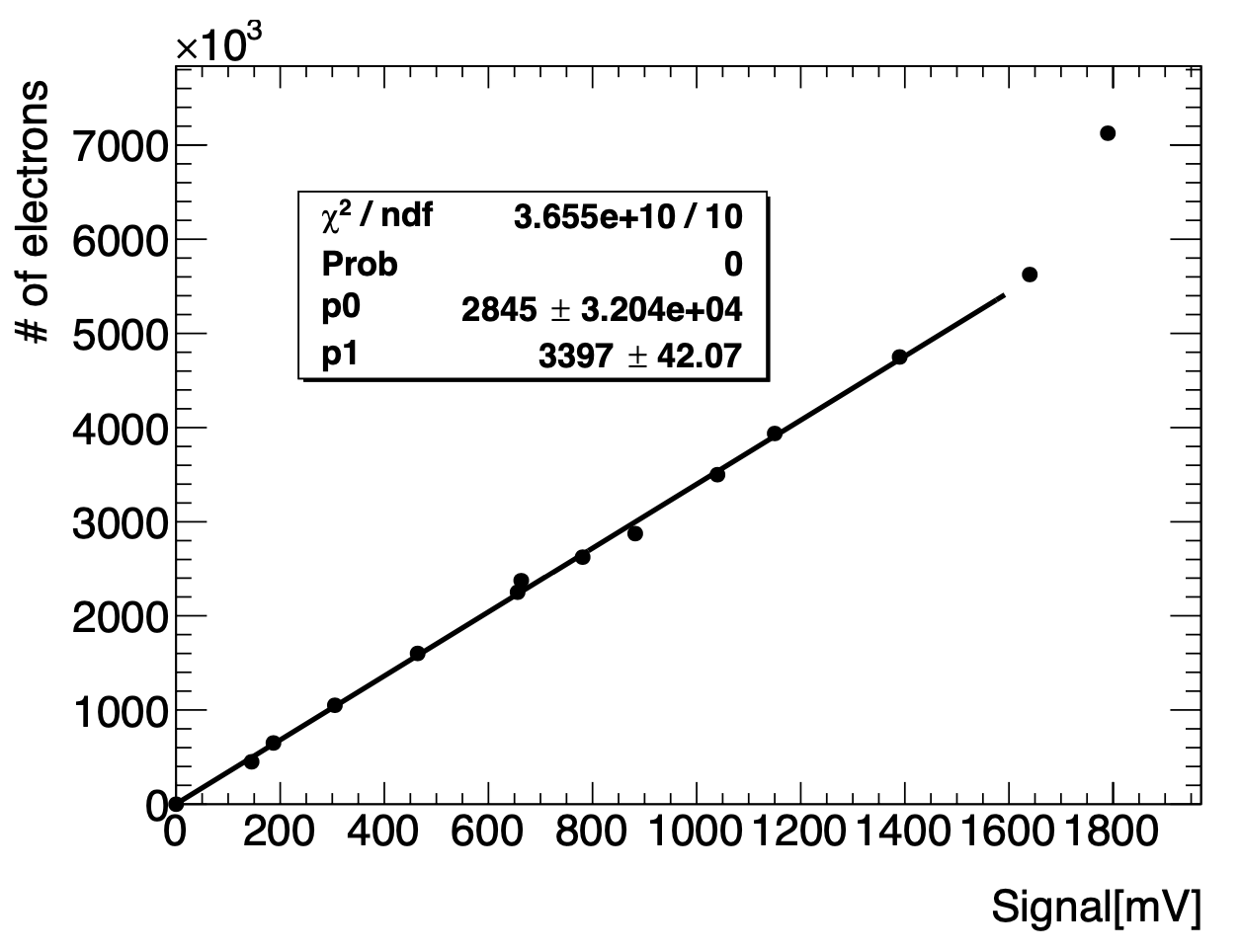}
         \caption{}
         \label{fig:cal_curve}
     \end{subfigure}
\caption{Calibration curve measurement a) Schematic for estimating calibration curve b) Calibration curve }
        \label{fig:cal_est_setup}
\end{figure}

The effective gain of the GEM detector is defined as the ratio of the number of
avalanche electrons and the number of primary electrons.
Experimentally, the effective gain  can be determined in two different ways. \\
\begin{itemize}
    \item \textbf{Method 1 :} Measuring induced current from the avalanche electrons at the readout board. One can then extract the effective gain of the detector by using the following formula\\
    \begin{equation}
    \label{gain_I}
  G_{{\rm eff}} = \frac{I_{{\rm}anode}}{e\times n_{{\rm primary}}\times R_{{\rm}x-ray}}
\end{equation}
where , \begin {itemize}
\item G$_{{\rm eff}}$ is the effective gain of the detector.
\item I$_{{\rm anode}}$ is the measured induced current from the readout board.
\item $e$ is the electron charge. 
\item \textcolor{black}{n$_{{\rm primary}}$ is the number of primary electrons in the drift volume of detector due to complete absorption of 5.9 KeV photons from $^{55}{\rm Fe}$ source. } 
\item R$_{{\rm x-ray}}$ is the rate of the $^{55}{\rm Fe}$ source.  
\end{itemize} 

It is to be noted that the denominator in Eqn.~\ref{gain_I} corresponds to the
induced current due to primary electrons at the drift cathode. I$_{{\rm anode}}$ was
measured by summing all the pads of the readout board and using a highly 
sensitive picoammeter. A collimated $^{55}{\rm Fe}$ source was used so R$_{{\rm x-ray}}$ was
estimated by noting the \textcolor{black}{count} rate of the $^{55}{\rm Fe}$ spectra from the detector. \\
    
 \item \textbf{Method 2 :} By extracting a monochromatic source spectra from the bottom of the last GEM or
 readout board and extracting the number of avalanche electrons embedded
 inside the spectra. The ratio of the number of avalanche electrons and
 the number of primary electrons from ionization of the gas inside the
 drift region because of the passage of primary ionizing particles will be
 the effective gain of the detector. In this method, effective gain can be formulated as follows 
 \begin{equation}
    \label{gain_spec}
  G_{{\rm eff}} = \frac{N_{{\rm avalanche}}^{{\rm spectra}}}{n_{{\rm primary}}}
\end{equation}
where \begin {itemize}
\item G$_{{\rm eff}}$ is the effective gain of the detector.
\item N$_{{\rm avalanche}}^{{\rm spectra}}$ is the number of avalanche electrons in the monochromatic source spectra from GEM detector.
\item \textcolor{black}{n$_{{\rm primary}}$ is the number of primary electrons in the drift volume of detector due to complete absorption of 5.9 KeV photons from $^{55}{\rm Fe}$ source. } 
\end{itemize}
 The total number of avalanche
 electrons was estimated by  first measuring the correlation between
 the known amount of input charge via high precission 1 pF capacitor to the readout electronics and the corresponding
  signal from the readout electronics. One can correlate this 1 pF capacitor to GEM detector. The schematic for such a measurement is shown in Fig.
 ~\ref{fig:cal_scheme}. The resulting calibration curve is shown in
 Fig.~\ref{fig:cal_curve}. It is clear from the calibration curve that
 the electronics in the experimental set up reaches saturation region when
 the signal goes above 1.6 V, which is fairly common in
 electronics. So, care must be taken while operating the detector
 to ensure the electronics don't reaches the saturation region in
 order to avoid an incorrect measurement of the effective gain. 
\end{itemize}
It is expected that the measurements of effective gain from the
methods  described above will provide different results because of the
mismatch in the number of avalanche electrons that reach the bottom
of the last GEM due to the bending of electric field lines and the avalanche
electrons reaching the readout board \textcolor{black}{which causes variation in electron transparency~\cite{GEM_sauli2}}. The difference in the number of
avalanche electrons reaching the readout board and accumulating
at the bottom of the last GEM is largely dependent on the electric field 
between the last GEM and the readout board (induction gap field).

This can be shown by effective gain measurements using both  methods
for different fields across the induction gap. This measurement was done by using voltage divider circuit to bias each electrode of the triple
GEM detector, which uses a single channel of the high voltage power supply. Varying the output voltage from the power supply provided different
fields across the induction gap and also potential differences across
each GEM layer. The resistors in the voltage divider circuit were chosen so
that the variation of potential difference across each GEM during the
variation of the output voltage from the power supply was the same. Table~\ref{table:1} shows
the output voltage from the power supply
and the corresponding potential difference across each GEM and the
induction field.
\begin{table}[h!]
\centering
 \begin{tabular}{||c | c |c|c| c ||} 
 \hline
 Input voltage to  & Potential difference across  & TG1 [kV/cm]  & TG2 [kV/cm] & Induction  \\ 
 voltage divider [V]  &  each GEM [Volt] &   &   &  field [kV/cm] \\ [0.5ex]
 \hline\hline
 -3550 & 355 & 2.23  & 2.23 & 1.7\\ 
 \hline
 -3700 & 370 & 2.32 & 2.32  & 1.9 \\
 \hline
 -3750 & 375 & 2.36 &  2.36 & 2.0 \\ [1ex]
 \hline
 \end{tabular}
 \caption{Table for input voltage to voltage divider and corresponding potential difference across GEMs and induction field.}
\label{table:1}
\end{table}

\begin{figure}[h!]
     \centering
     \begin{subfigure}[h]{0.3\textwidth}
         \centering
         \includegraphics[width=\textwidth]{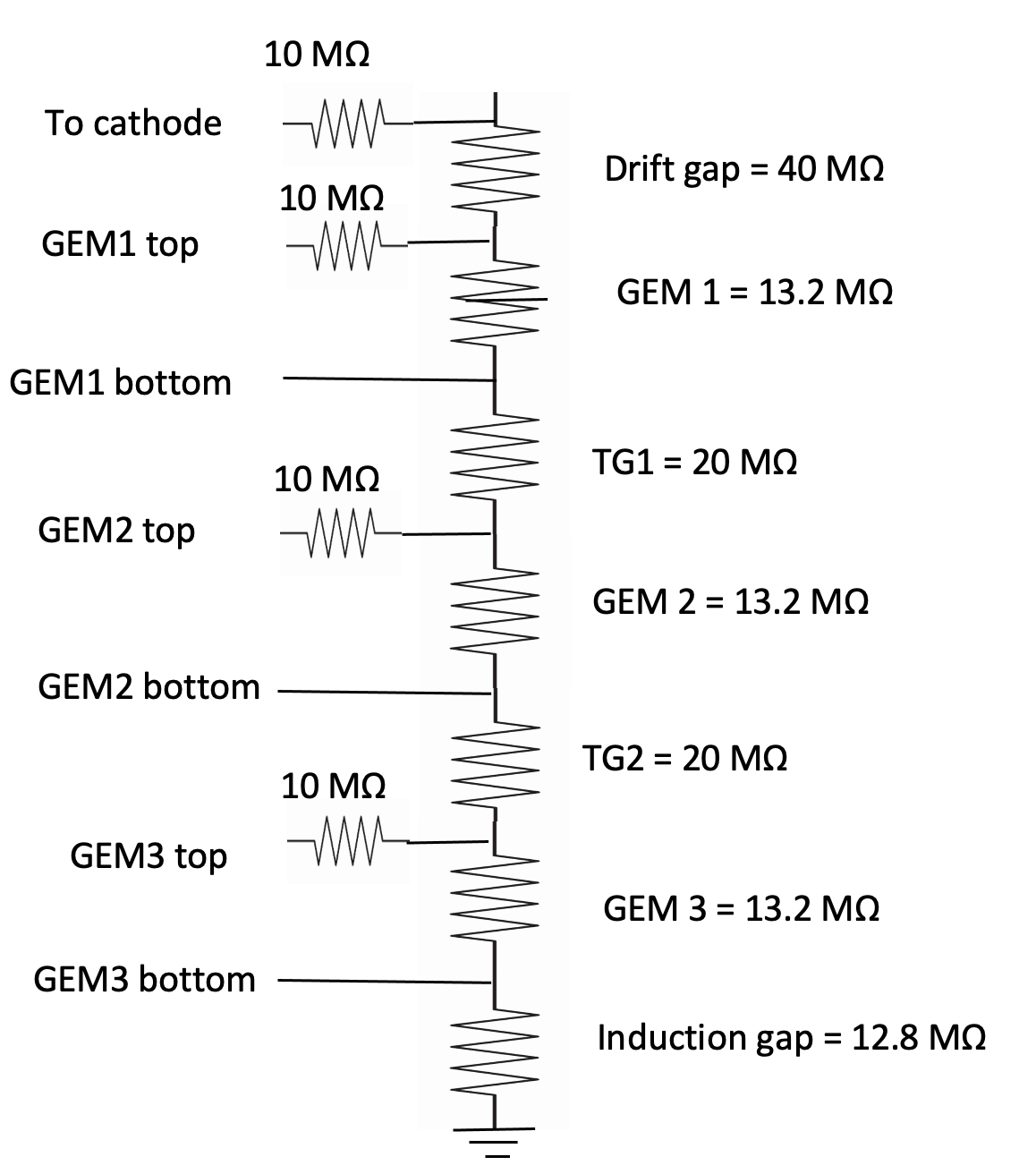}
         \caption{}
         \label{fig:potdivider}
     \end{subfigure}
     \begin{subfigure}[h]{0.3\textwidth}
         \centering
         \includegraphics[width=\textwidth]{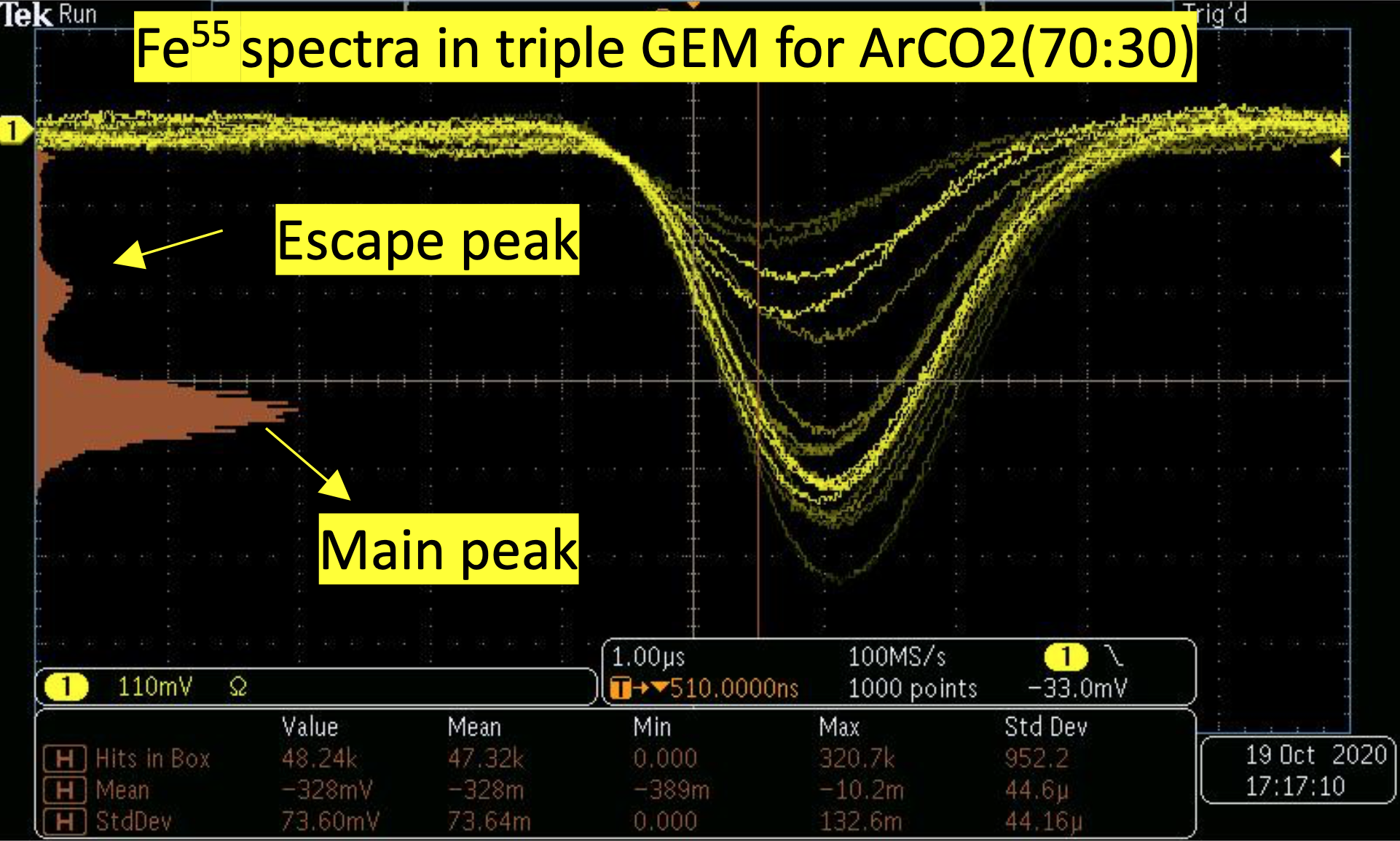}
         \caption{}
         \label{fig:fe55spectra}
     \end{subfigure}
     \hfill
     \begin{subfigure}[h]{0.3\textwidth}
         \centering
         \includegraphics[width=5.3 cm, height = 5.0 cm]{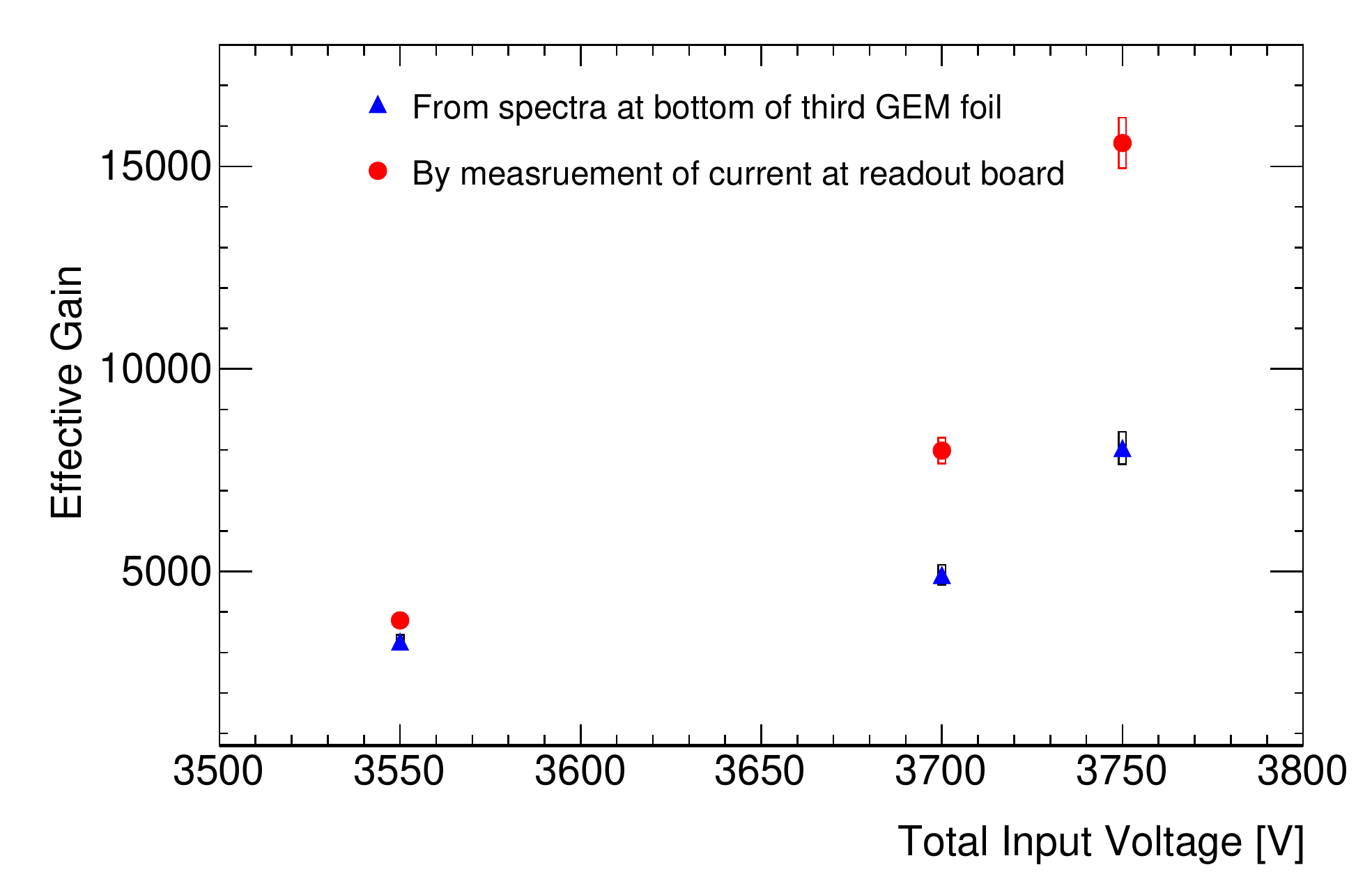}
         \caption{}
         \label{fig:gain_curr_spectra}
     \end{subfigure}
\caption{Signal from triple GEM detector \textcolor{black}{a) Voltage divider conifguration for this measurement} b) $^{55}{\rm Fe}$ spectra from triple GEM detector in Ar:CO$_{2}$ (70:30) gas mixture c) Effective gain measurement from the two methods described in section~\ref{subsec:effgain}.\textcolor{black}{Systematic error of 5$\%$ is assigned to red data points which is based on  calibration curve and spectra measurement as per description in the text. A systematic error of 3$\%$ is assigned to blue data point which is based on anode current measurement. The 4$\%$ uncertainty involves 2$\%$ uncertainty in anode current due to noise , 2$\%$ uncertainty in gas mixing unit and 2$\%$ uncertianty in voltage divider circuit. The statistical errors are rather small and are of the order of size of markers. }  }
        \label{fig: gain_spectra}
\end{figure}

\begin{figure}[h!]
 \centering
 \includegraphics[width=8. cm, height = 6.0 cm]{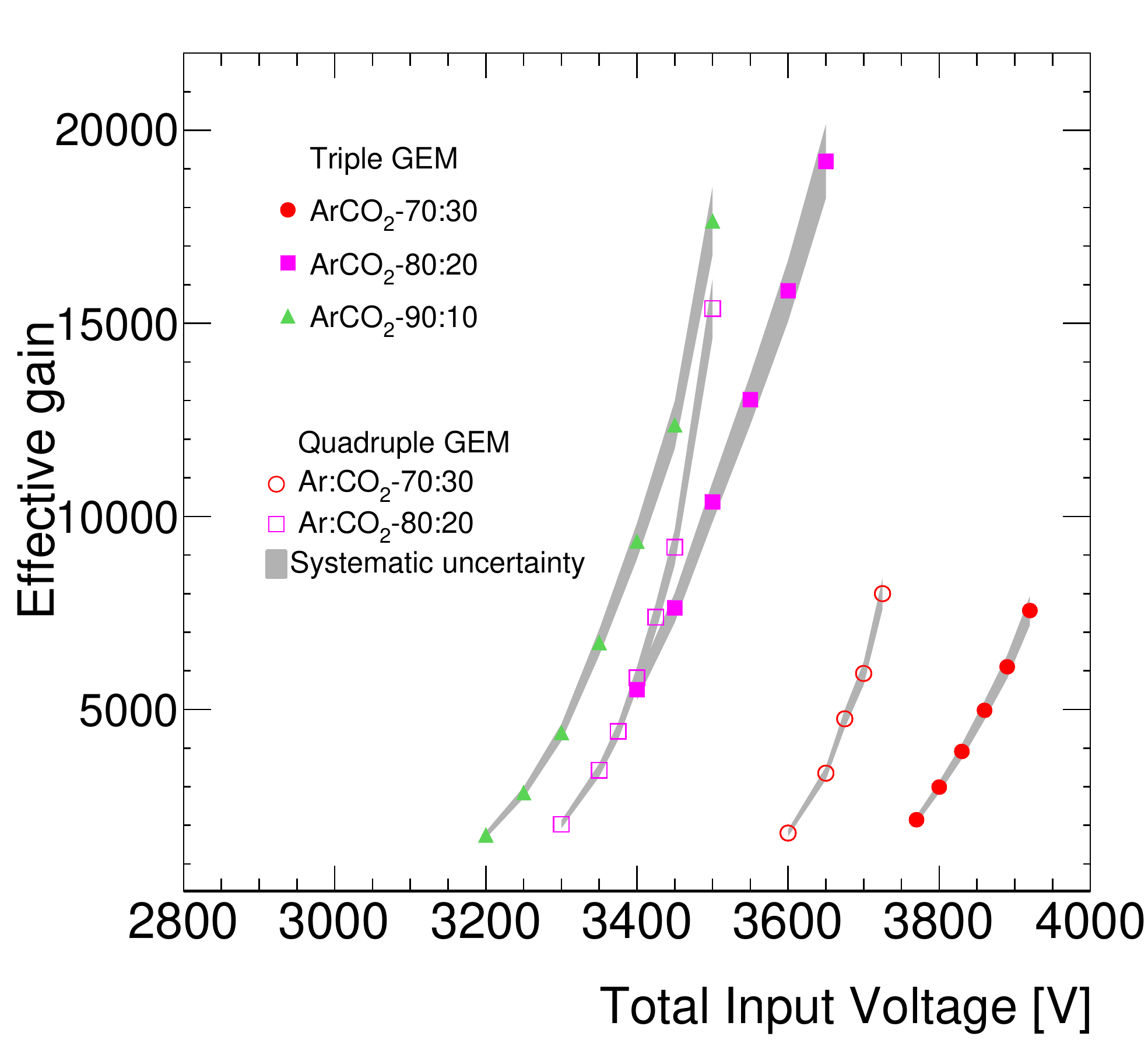}
 \caption{Effective gain from triple and quadruple GEM detectors for various composition of Ar:CO$_{2}$ gas mixtures. \textcolor{black}{The systematic error of 5$\%$ is shown by the gray band while statistical errors are of the order of size of markers.}}
 \label{fig:gain_scan_argas}
 \end{figure}

Fig.~\ref{fig:fe55spectra} shows $^{55}{\rm Fe}$ spectra in the triple GEM detector
using the Ar:CO$_{2}$ (70:30) gas mixture for \textcolor{black}{voltage of -3700 V to  the detector via potential divider} while Fig.~\ref{fig:gain_curr_spectra} shows the effective gain calculation
from both the methods described earlier. \textcolor{black}{The systematic error on effective gain is estimated to be 5$\%$ by taking into account of uncertainty of 2$\%$ in gas mixing unit, 4$\%$ tolerance in capactor of calibration set up and 2$\%$ uncertainty in output voltage from voltage divider.} One can clearly see that, as
the induction field becomes larger, more avalanche electrons 
reach the readout board. which in turn increases the effective gain
with respect to that estimated by using the spectra from the bottom of
the last GEM. \textcolor{black}{Also at higher applied voltage the effective gain measured from anode current increases at faster rate as compared to the calculation where signal from bottom of last GEM foil was used. This is because effective gain is an exponential function of voltage across the GEM foil~\cite{GEM_sauli} and is also a function of electron extraction efficiency in each of the transfer gaps and induction gap. At large input voltage to the GEM detector via voltage divider , the gap fields increases and also potential difference acorss each GEM foil. Both these factors causes exponential increase in avalanche electrons reaching the anode and hence larger difference in effective gain at higher input voltage from voltage divider to the triple GEM detector.}
 Effective gain scanning for both quadruple and triple GEM detectors were performed for various argon-based gas mixtures \textcolor{black}{and is shown in fig.~\ref{fig:gain_scan_argas}}. The study was done using a voltage divider scheme for both triple and 
 quadruple GEM detectors. For a higher argon content in the gas mixture , one can
 attain a greater effective gain at the same operating voltage. This is because, since argon is the 
 primary ionising gas component, there will be larger number of primaries with the higher argon fraction.
 Also, the quadruple GEM detector will attain the same effective gain at a lower operating voltage
 as compared to the triple GEM detector using the same gas mixture because of the additional avalanche layer.

\subsubsection{Ion Back Flow}
\begin{figure}[h!]
     \centering
     
     \begin{subfigure}[h]{0.5\textwidth}
         \centering
       \includegraphics[width=7. cm, height = 4.8 cm]{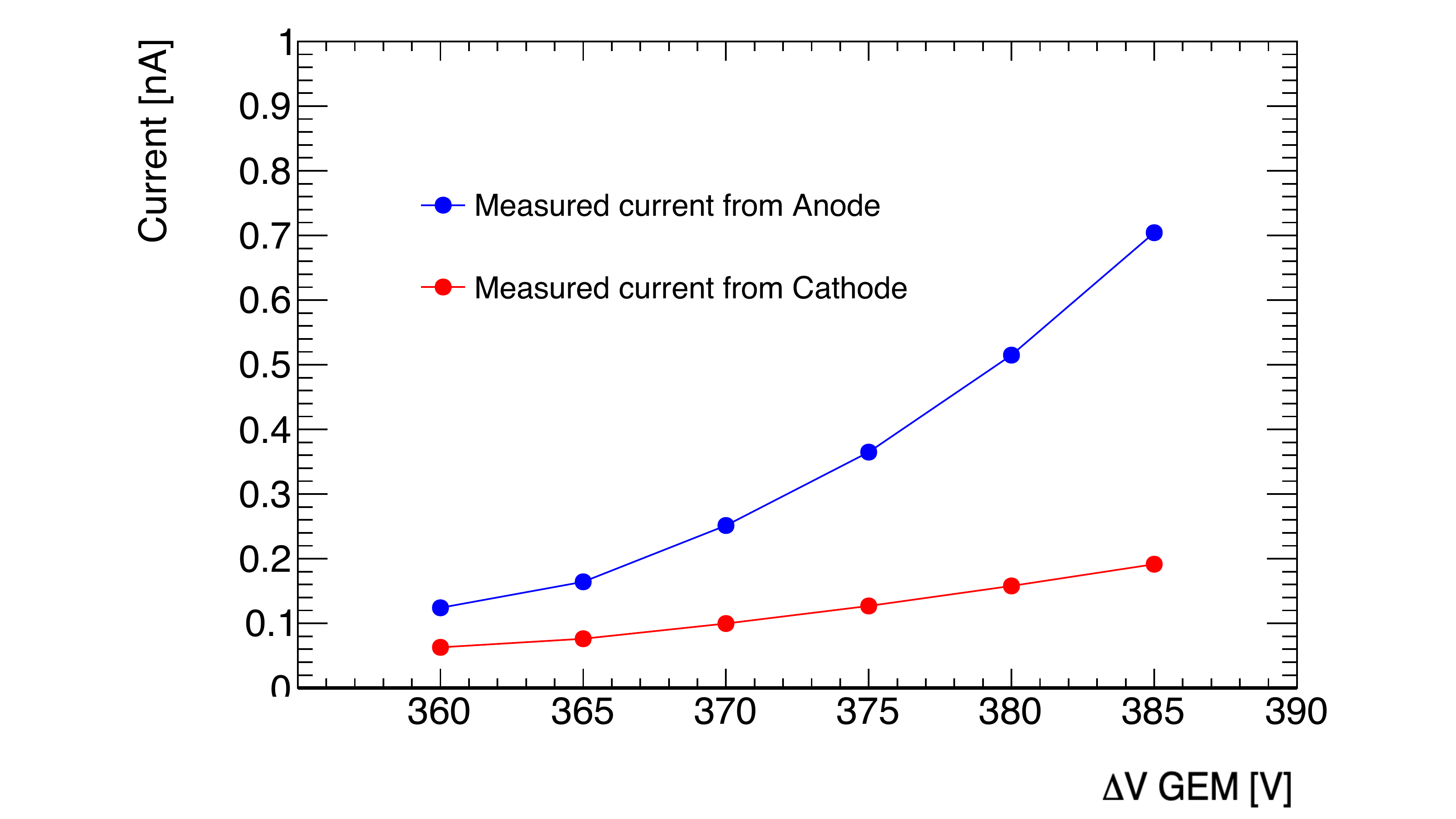}
         \caption{}
         \label{fig:tripGEM_current}
     \end{subfigure}
     \hfill
     \begin{subfigure}[h]{0.4\textwidth}
         \centering
         \includegraphics[width=7.0 cm, height = 4.8 cm]{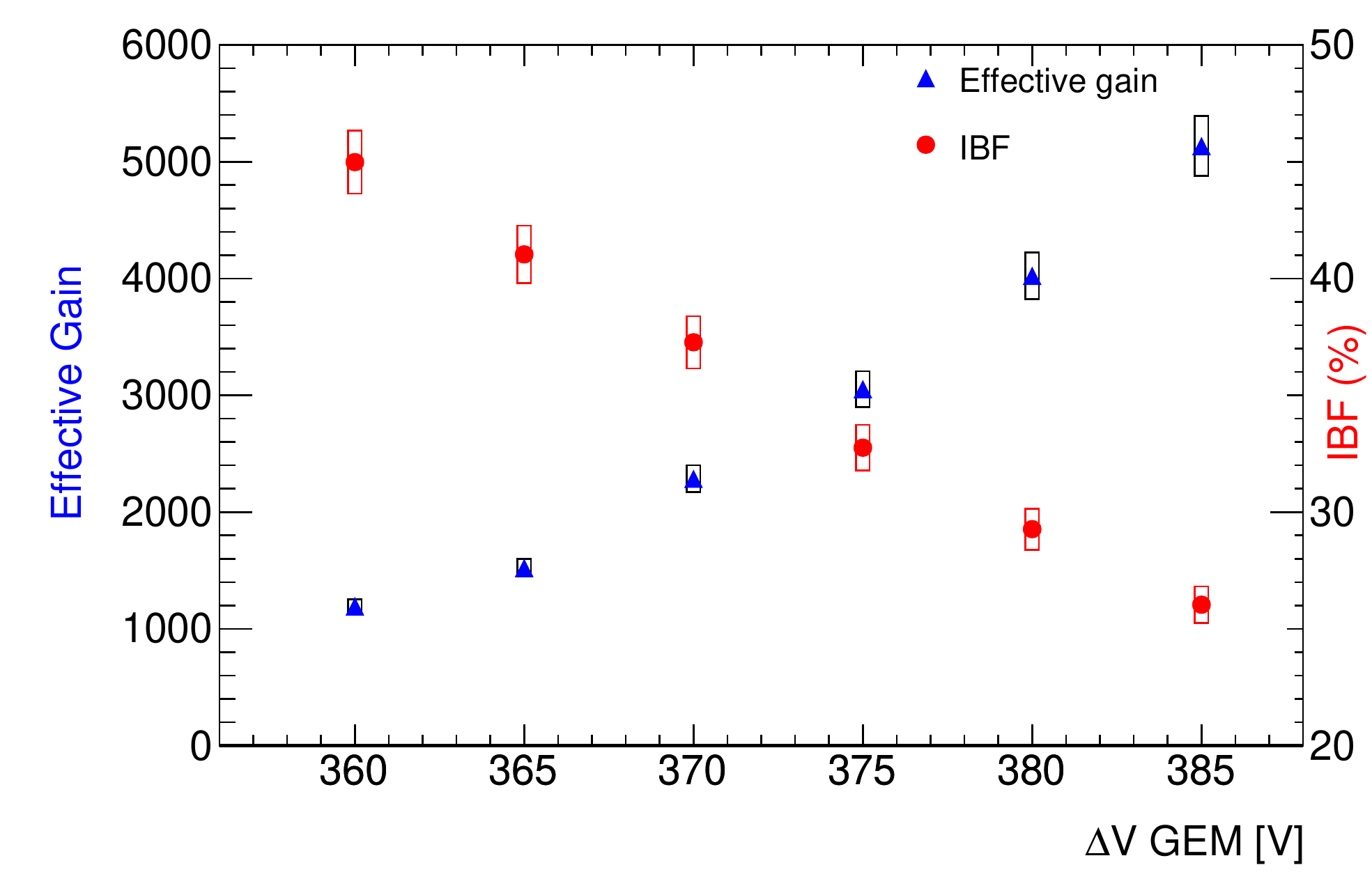}
         \caption{}
         \label{fig:tripGEM_gain_ibf}
     \end{subfigure}
\caption{Triple GEM detector IBF and effective gain measurement in Ar:CO$_{2}$ (70:30) gas mixture a) Current measured from anode and cathode during operation of X-ray tube.\textcolor{black}{ Both statistical and systematic errors are small and are of the order of the size of markers.} b) Effective gain and IBF measurement for different potential differences across the GEM layers. \textcolor{black}{The systematic errors are shown by rectangular box around the data points. The statistical errors are small and are of the order of the size of markers.} }
        \label{fig: tripGEM_XrayI_IBF_gain}
\end{figure}

\begin{figure}[h!]
     \centering
     
     \begin{subfigure}[h]{0.5\textwidth}
         \centering
       \includegraphics[width=7. cm, height = 4.8 cm]{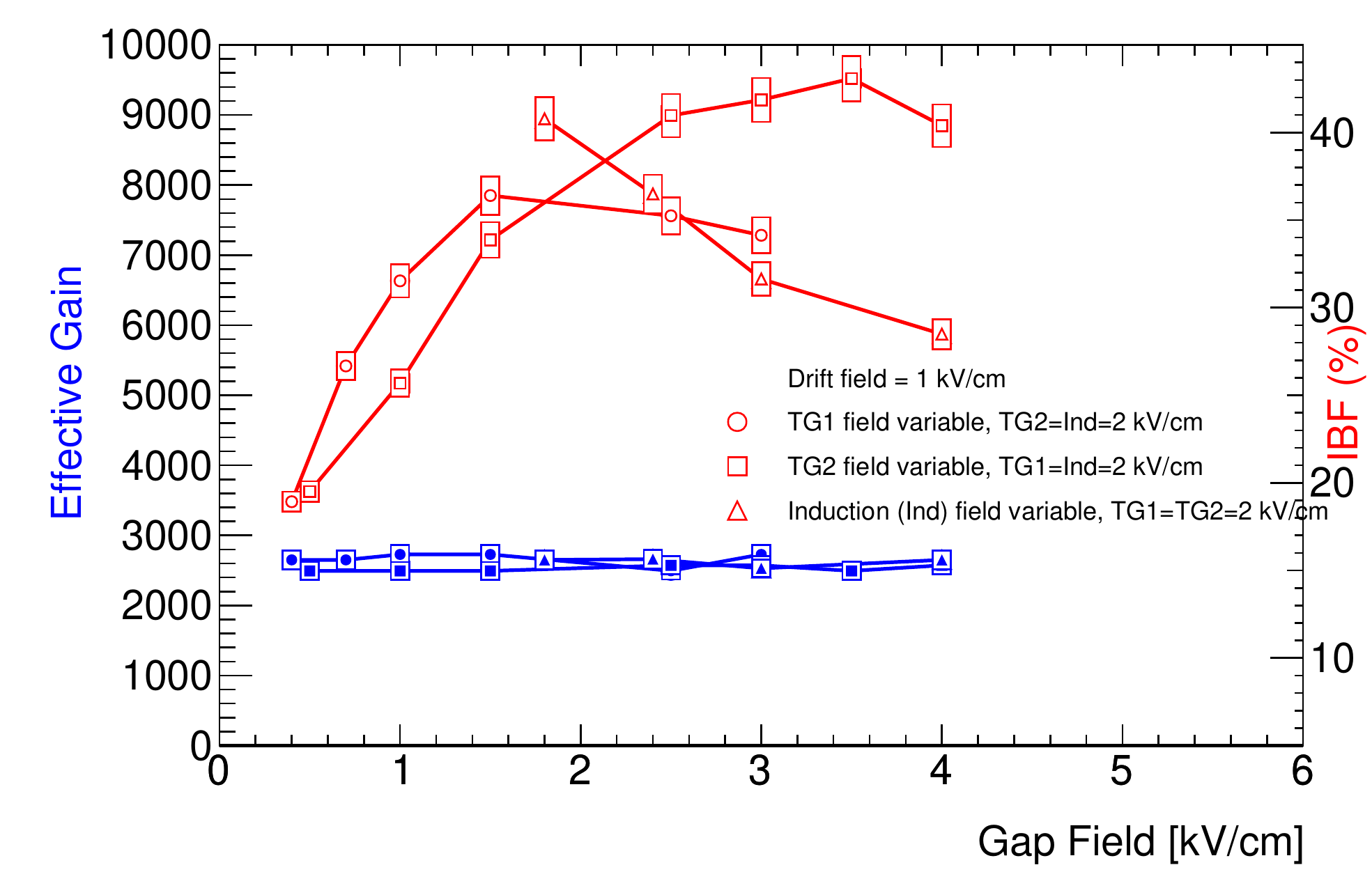}
         \caption{}
         \label{fig:tripGEM_ibf_gapfield}
     \end{subfigure}
     \hfill
     \begin{subfigure}[h]{0.4\textwidth}
         \centering
         \includegraphics[width=7.0 cm, height = 4.8 cm]{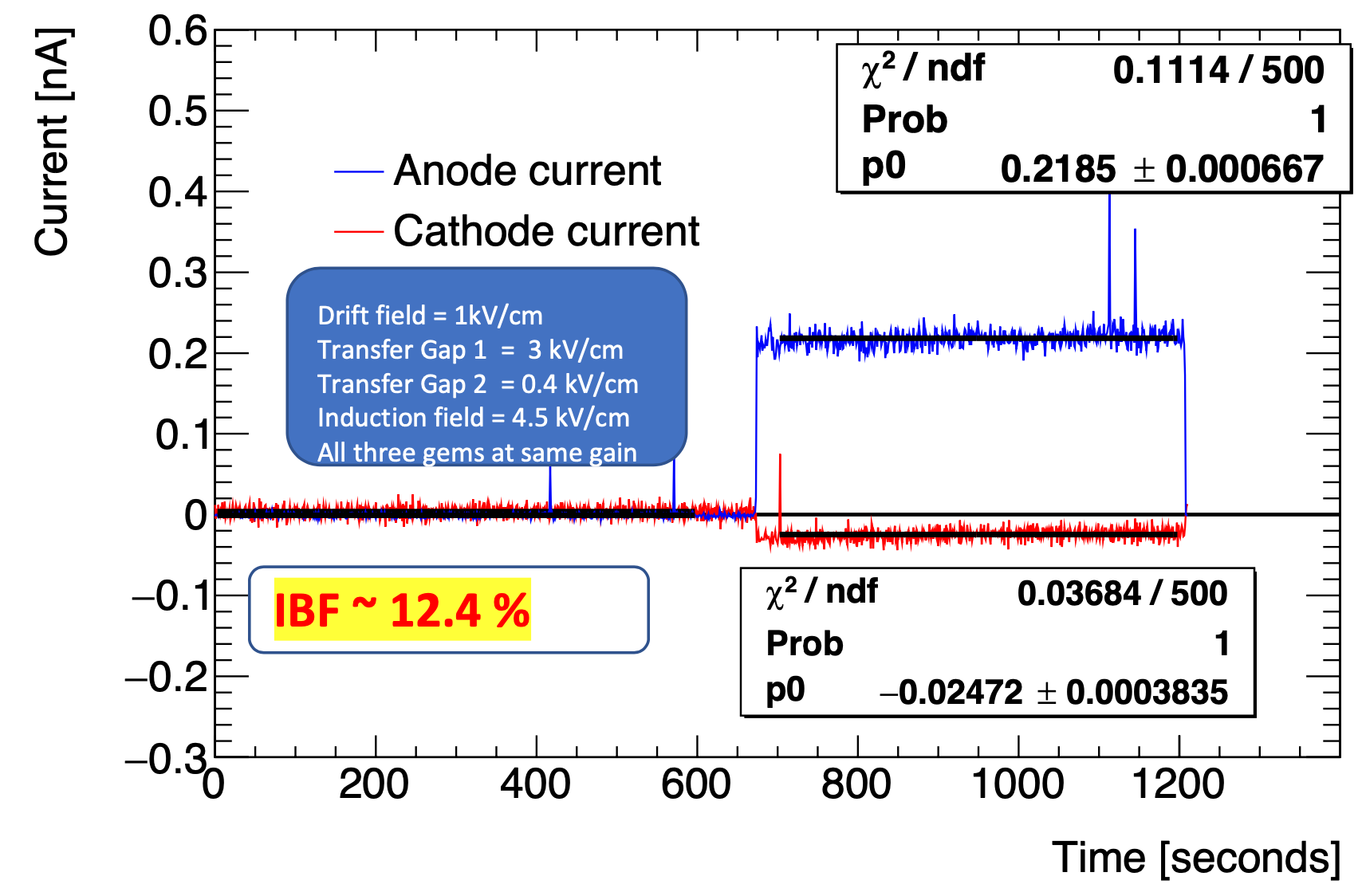}
         \caption{}
         \label{fig:tripGEM_lowest_ibf_I}
     \end{subfigure}
\caption{Triple GEM detector IBF  measurement in Ar:CO$_{2}$ (70:30) gas mixture a) IBF estimated by varying a specific gap field while keeping other gap fields constant  b) Measured anode and cathode currents while operating the triple GEM detector with gap fields selected to reduce IBF. }
        \label{fig: tripGEM_ibf_optimize}
\end{figure}

Ion Back Flow (IBF) is the fraction of positive ions drifting towards to the drift region of 
GEM detector. Experimentally it is determined by taking the ratio of measured current from cathode and from anode~\cite{MORMANN2004315}.

\begin{equation}
    \label{IBF_eqn}
  IBF = \frac{I_{{\rm cathode}}}{I_{{\rm anode}}}
\end{equation}
For the measurement and minimization of IBF, both triple and quadruple GEM detector electrodes were
biased by individual channels of a multichannel power supply as shown in
Fig.~\ref{fig:triple_GEM} and in Fig.~\ref{fig:quad_GEM}. Only Ar:CO$_2$ 70:30 gas was used for detailed study of IBF.  The detector drift area was irradiated
with a Ag target X-ray tube  fitted with a collimator \textcolor{black}{of 1mm diameter}. The X-ray tube was operated at 20 kV
peak voltage and 18 uA current.  During IBF estimation for both triple and quadruple GEM detector, all the layers of the GEM were operated at the same potential difference. 
Fig.~\ref{fig:tripGEM_current} shows the current measured from the anode and cathode of triple GEM
detector after turning on the X-ray tube for different potential differences across the GEM \textcolor{black}{foil}. The measurements were done by keeping the drift field at 1 kV/cm while the transfer gap and induction fields were at 2 kV/cm.
Using equation ~\ref{IBF_eqn} and results from Fig.~\ref{fig:tripGEM_current} , IBF was
estimated for different potential differences across the GEM as shown in Fig.~\ref{fig:tripGEM_gain_ibf}. \textcolor{black}{The IBF is assigned a systematic uncertainty of 3$\%$ by taking into account of 2$\%$ systematic uncertainty on each of the anode and cathode current measurement due to noise and 1$\%$ uncertainty in the fitting procedure to anode and cathode current data. Further the noise in anode and cathode current was estimated by measuring them for long time before biasing the GEM detector and after turning off the X-Ray tube.} At the same time, effective gain for different voltages across the GEM \textcolor{black}{foils} was measured \textcolor{black}{and the} results are shown in Fig.~\ref{fig:tripGEM_gain_ibf}. 
\newline
During long-term operation of GEM-based detectors in particle physics experiments \textcolor{black}{like TPC in ALICE ~\cite{GARABATOS2004197,KETZER2013237} or HBD in PHENIX ~\cite{AIDALA2003200}}, the GEMs are
operated at the same effective gain. In GEM-based tracking detectors \textcolor{black}{like TPC or recently proposed Transition Radiation Detector (TRD)~\cite{BARBOSA2019162356}} where maintaining uniform
electric field is important for determining accurate trajectories of particles, minimizing IBF is
a major priority\textcolor{black}{~\cite{GARABATOS2004197,KETZER2013237,SAULI2006269}}. Large IBF tends to cause a distortion of the uniform electric field because of the slow drift velocity of heavier positive ions. \textcolor{black}{Over the past deccade several research groups has worked significantly to reduce the IBF both without the effect of magnetic field ~\cite{Luz_2018,Ball_2014,SAULI2006269,MORMANN2004315} and also under the effect of magenetic field~\cite{KILLENBERG2004251}.} One way to minimize IBF for a specific effective gain of the
GEM detector is to operate the detector such that the electric fields in different gaps are not
the same ~\cite{KILLENBERG2004251,Ball_2014,SAULI2006269,Luz_2018}. In this study for minimizing IBF in triple GEM detector, an effective gain of 4500 was chosen and the corresponding IBF as in Fig~\ref{fig:tripGEM_gain_ibf} was 36$\%$. To see the 
effect of different gap fields on IBF for the same effective gain, the triple GEM detector was
operated by varying a specific gap field while keeping other gap fields constant and
maintaining the same effective gain by changing the potential difference across the GEMs. All
the GEMs were kept at the same potential difference to avoid the effect of ion blocking because of having a
different gain of each GEM. \textcolor{black} {It has been shown explicitely that operating each GEM foil at different gain compared to the other foil in the same GEM detector will cause IBF suppression.~\cite{MORMANN2004315} . Operating the triple GEM detector with symmetric powering of GEM foils and only changing the gap fields will allow us to understand only the effect of gap field on IBF suppression.} Fig.~\ref{fig:tripGEM_ibf_gapfield} shows the estimated IBF as a function of a specific gap field while keeping the other gap fields constant. \textcolor{black}{The study of asymmetric variation of electric field in different gap fields on IBF was done in three steps as below }.
\begin{itemize}
\item \textcolor{black}{Varying trasfer gap 1 (TG1) field while keeping transfer gap 2 (TG2) and induction field constant at $\sim$ 2 kV/cm. } 
\item \textcolor{black}{Varying TG2 field while keeping TG1 and induction field constant at $\sim$ 2 kV/cm.}
\item \textcolor{black}{Varying induction field while keeping TG1 and TG2 fields constant at $\sim$ 2 kV/cm.}
\end{itemize}

\begin{figure}[h!]
     \centering
     \begin{subfigure}[h]{0.5\textwidth}
         \centering
       \includegraphics[width=7. cm, height = 4.5 cm]{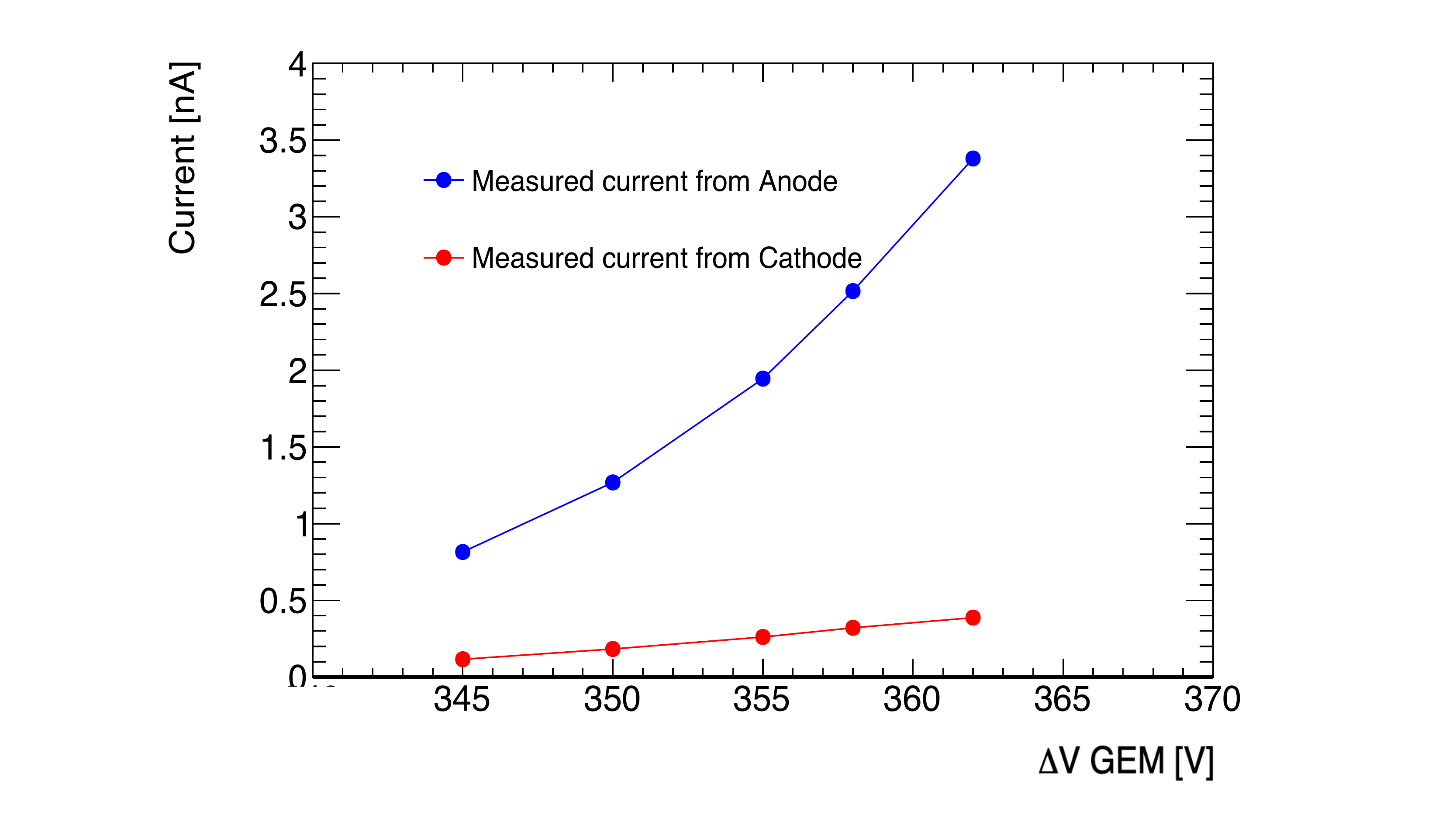}
         \caption{}
         \label{fig:quadGEM_current}
     \end{subfigure}
     \hfill
     \begin{subfigure}[h]{0.4\textwidth}
         \centering
         \includegraphics[width=7.0 cm, height = 4.5 cm]{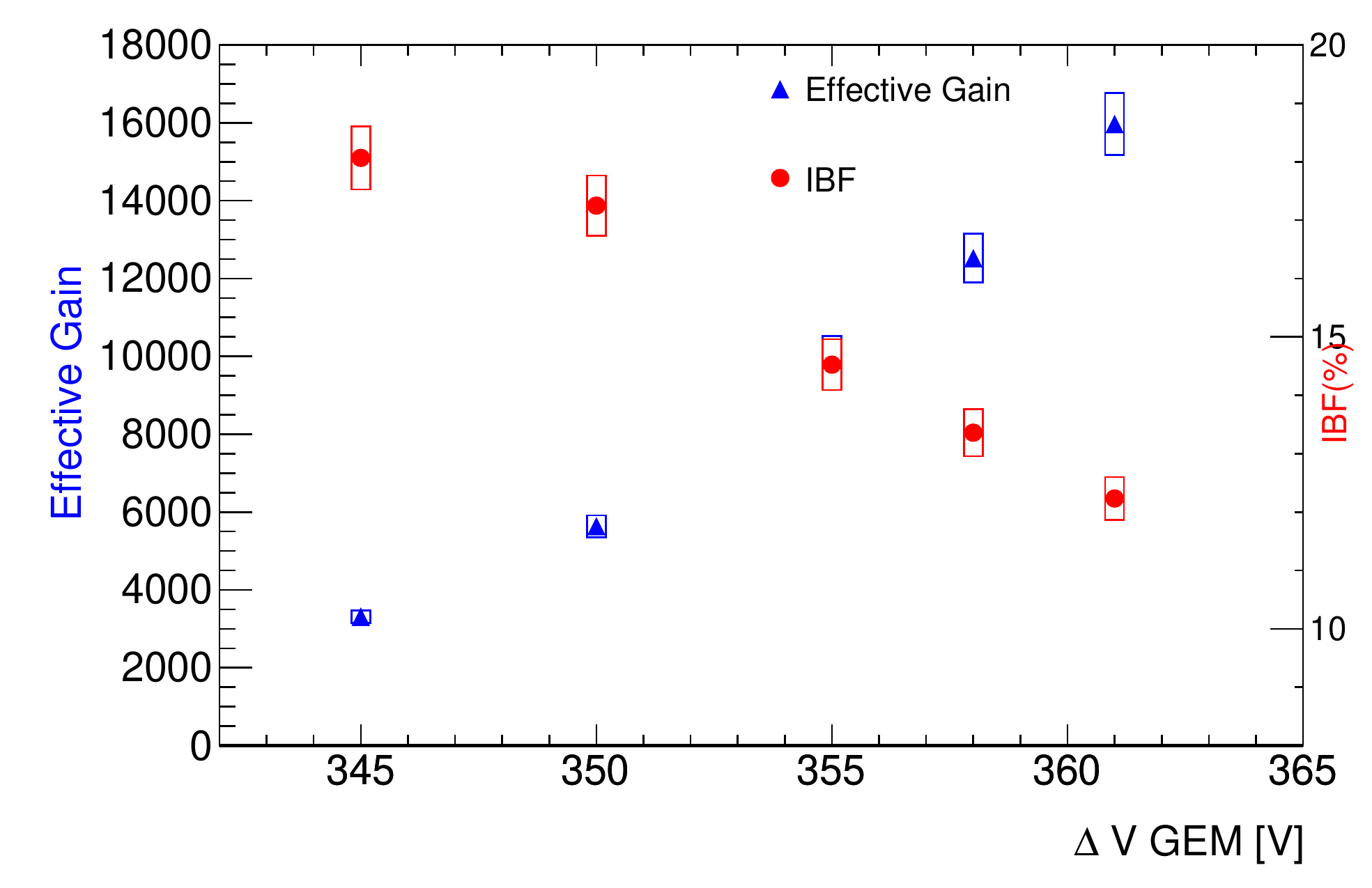}
         \caption{}
         \label{fig:quadGEM_gain_ibf}
     \end{subfigure}
\caption{Quadruple GEM detector IBF and effective gain measurement in Ar:CO$_{2}$ (70:30) gas mixture a) Current measured from anode and cathode during operation of X-ray tube. \textcolor{black}{Both of them has systematic error of 2$\%$ due to noise. Boith statistical and systematic errors are small and are of the same order as markers.} b) Effective gain and IBF measurement for different potential differences across the GEM layers. \textcolor{black}{The systematic errors are shown by rectangular box around the data points. The statistical errors are small and are of the order of the size of markers.} }
        \label{fig: quadGEM_XrayI_IBF_gain}
\end{figure}

\begin{figure}[h!]
     \centering
     
     \begin{subfigure}[h]{0.5\textwidth}
         \centering
       \includegraphics[width=7. cm, height = 4.8 cm]{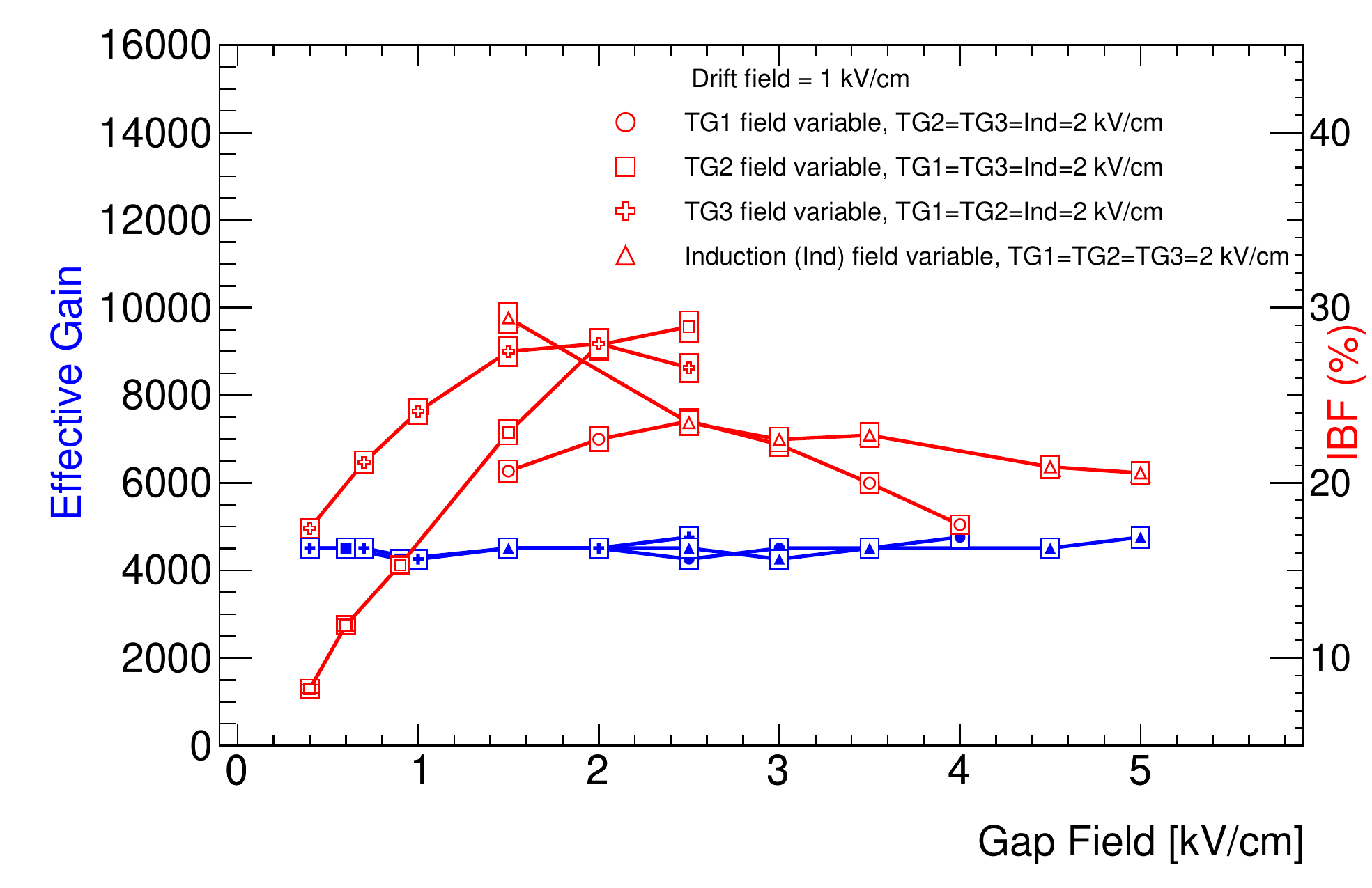}
         \caption{}
         \label{fig:quadGEM_ibf_gapfield}
     \end{subfigure}
     \hfill
     \begin{subfigure}[h]{0.4\textwidth}
         \centering
         \includegraphics[width=7.0 cm, height = 4.8 cm]{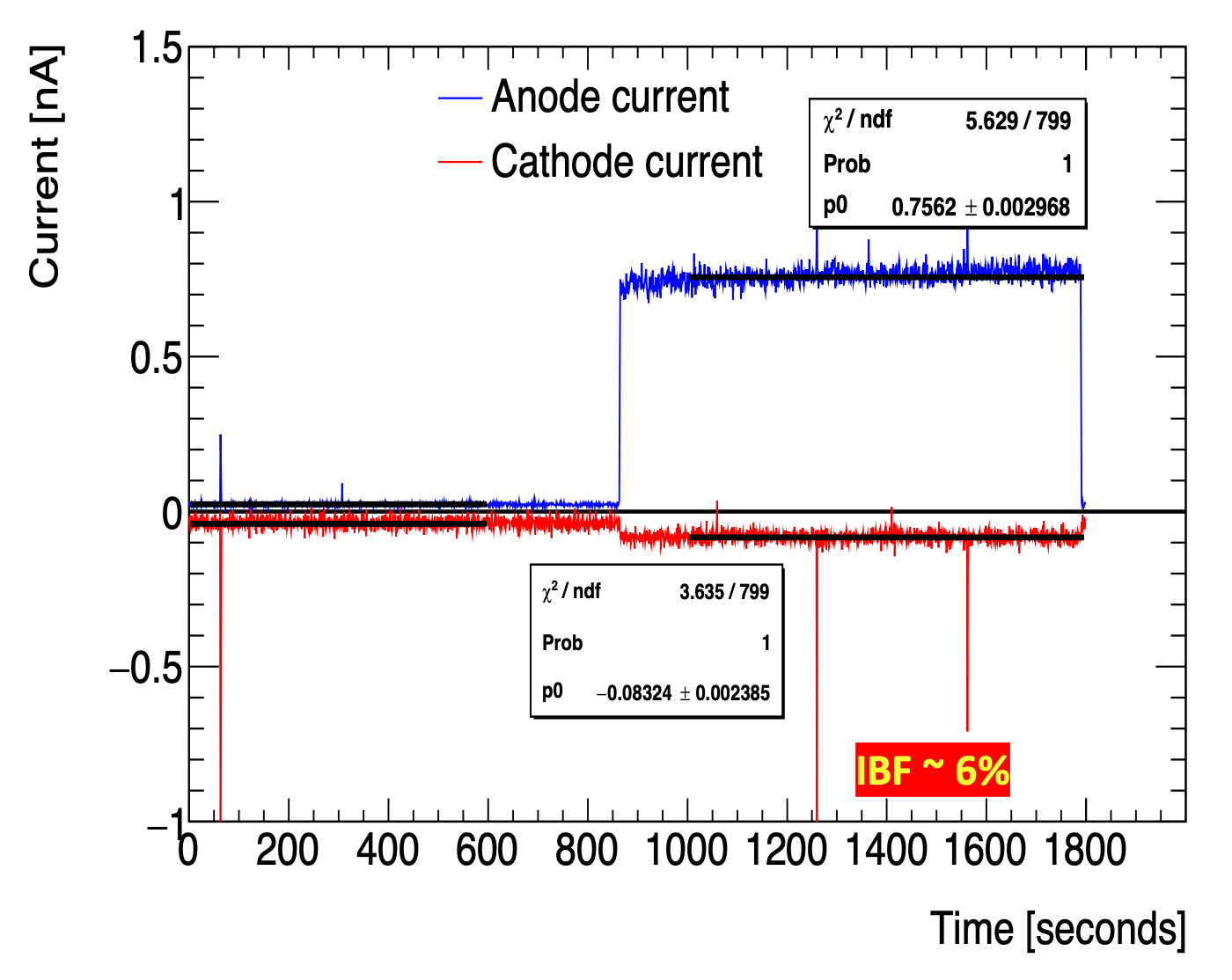}
         \caption{}
         \label{fig:quadGEM_lowest_ibf_I}
     \end{subfigure}
\caption{Quadruple GEM detector IBF  measurement in Ar:CO$_{2}$ (70:30) gas mixture a) IBF estimated by varying a specific gap field while keeping other gap fields constant.\textcolor{black}{The systematic errors are shown by rectangular box around the data points. The statistical errors are small and are of the order of the size of markers.}  b) Measured anode and cathode currents while operating the quadruple GEM detector with gap fields chosen to reduce IBF. }
        \label{fig: quadGEM_ibf_optimize}
\end{figure}

The effective gain was kept constant as shown in Fig.~\ref{fig:tripGEM_ibf_gapfield}. It can be seen that if transfer gap 2 and transfer gap 1 fields are kept low then one can reduce IBF. However if transfer gap 2 is kept at a lower electric field than transfer gap 1 field, then IBF suppresion is greater. Also, keeping the induction field at a higher value suppresses IBF. Keeping these parameters in mind, the triple GEM detector was operated so that transfer gap 1 and induction gap fields were large, while the transfer gap 2 field was kept low. The measured anode and cathode currents are shown in Fig.~\ref{fig:tripGEM_lowest_ibf_I} and using equation ~\ref{IBF_eqn} the estimated IBF for this voltage configuration is about 12.4$\%$, which is significantly lower than the value obtained with effective gain of 4500. 
\newline
A similar study was done with the quadruple GEM detector. First, the quadruple GEM detector was scanned for different effective gain by changing the potential difference across the GEMs, which is shown in Fig.~\ref{fig:quadGEM_gain_ibf} and, at the same time, IBF was estimated for that specific effective gain by using the X-ray tube as the ionizing source and measuring the anode and cathode current from the quadruple GEM detector as shown in Fig. ~\ref{fig:quadGEM_current}.
\newline
In order to study the effect of gap fields and optimizing them for reducing the IBF , an effective
gain of 4000 and corresponding IBF of 17$\%$ as per Fig.~\ref{fig:quadGEM_gain_ibf} were chosen.
The quadruple GEM detector was then operated by varying a specific gap field while keeping the
other gap fields constant. \textcolor{black}{The study of asymmetric variation of electric field in different gap fields on IBF was done in four steps as below }.
\begin{itemize}
\item \textcolor{black}{Varying trasfer gap 1 (TG1) field while keeping transfer gap 2 (TG2),  transfer gap 3 (TG3) and induction field constant at $\sim$ 2 kV/cm. } 
\item \textcolor{black}{Varying TG2 field while keeping TG1, TG3 and induction field constant at $\sim$ 2 kV/cm.}
\item \textcolor{black}{Varying TG3 field while keeping TG1, TG2 and induction field constant at $\sim$ 2 kV/cm.}
\item \textcolor{black}{Varying induction field while keeping TG1, TG2 and TG3 fields constant at $\sim$ 2 kV/cm.}
\end{itemize}
The effective gain throughout the variation of the gap field was kept
constant by manipulating the potential difference across each GEM foil. All the four GEM foils were
operated at the same potential difference \textcolor{black}{for the same reason as mentioned in triple gem detector studies}. Fig.~\ref{fig:quadGEM_ibf_gapfield} shows the IBF
estimated for the quadruple GEM detector by varying a specific gap field while keeping the other gap
fields constant. From the same figure, it can be seen that the effective gain was kept constant
during the whole measurement. It is quite clear from this estimate that operating the detector
with transfer gap 1 and induction field at larger electric field while keeping the transfer gap
2 and transfer gap 3 field low will provide the maximum suppression of IBF. With this configuration, the quadruple GEM detector was again operated to estimate IBF. Fig.~\ref{fig:quadGEM_lowest_ibf_I} shows the anode and cathode current measured from the detector and using Equation~\ref{IBF_eqn} the IBF was estimated to be about 6$\%$, which is almost 3 times lower than the initial value. \textcolor{black}{Comparing the results from this work which focussed only on effect of gap fields on IBF and the results from ~\cite{MORMANN2004315} , which focussed on effect of asymmetric gem foil biasing , it can be seen that gap field has larger effect on IBF suppression as compared to only asymmetric gem foil biasing. This has been also confirmed in ~\cite{TARAFDAR2022167460}.}
\newline
\section{Conclusions}
Studies of effective gain, IBF and reduction of IBF for both triple and quadruple GEM
detectors were performed using Argon-based gas mixtures. The effective gain was calculated using two different
methods and the variation of results is as expected. Also, effective gain scanning was performed
for different Argon-based gas mixtures for both triple and quadruple GEM detectors. Higher argon
content increases the effective gain at the same potential difference across the GEMs
which is attributed to the fact that more ionising component is added to the gas mixture.
Also, the quadruple GEM showed higher effective gain as compared to the triple GEM detector for the same
potential difference across the GEMs because of the additional avalanche layer. The results for IBF
shows that the IBF can be suppressed by varying the electric field in different gap regions of the
detector in such a way that the low electric field in \textcolor{black}{transfer gap 2 for both triple and quadruple GEM detector} acts as an ion blocker. \textcolor{black}{In addition quadruple GEM detector shows larger IBF suppression compared to triple GEM detector due to the presence of additonal transfer gap where lower electric field aids in additional suppression of IBF. The studies on IBF suppression in this technical report quantifies IBF suppression by changing specific gap field of standard GEMs from CERN ~\cite{GDD} while keeping other gap fields constant for both triple and quadruple GEM. }

\bibliographystyle{unsrt}  
\bibliography{references}  

\begin{thebibliography}{10}

\bibitem{GEM_sauli}
F.~Sauli.
\newblock {GEM}: A new concept for electron amplification in gas detectors.
\newblock {\em Nucl. Instrum. Meth. A}, 386(2):531--534, 1997.

\bibitem{AIDALA2003200}
C.~{Aidala et al.}
\newblock {A Hadron Blind Detector for PHENIX}.
\newblock {\em Nuclear Instruments and Methods in Physics Research Section A:
  Accelerators, Spectrometers, Detectors and Associated Equipment}, 502(1):200
  -- 204, 2003.
\newblock Experimental Techniques of Cherenkov Light Imaging. Proceedings of
  the Fourth International Workshop on Ring Imaging Cherenkov Detectors.

\bibitem{GARABATOS2004197}
C.~Garabatos.
\newblock {The ALICE TPC}.
\newblock {\em Nuclear Instruments and Methods in Physics Research Section A:
  Accelerators, Spectrometers, Detectors and Associated Equipment}, 535(1):197
  -- 200, 2004.
\newblock Proceedings of the 10th International Vienna Conference on
  Instrumentation.

\bibitem{Tarafdar:2019oB}
Sourav Tarafdar.
\newblock {sPHENIX TPC simulation studies}.
\newblock {\em PoS}, MPGD2017:067, 2019.

\bibitem{GEM_sauli2}
Fabio Sauli.
\newblock {The gas electron multiplier (GEM): Operating principles and
  applications}.
\newblock {\em Nuclear Instruments and Methods in Physics Research Section A:
  Accelerators, Spectrometers, Detectors and Associated Equipment}, 805:2 --
  24, 2016.
\newblock Special Issue in memory of Glenn F. Knoll.

\bibitem{GDD}
https://gdd.web.cern.ch.
\newblock {\em Gas Detectors Development Group}.

\bibitem{ORTEC_PA}
https://www.ortec-online.com/products/electronics/preamplifiers/142a-b-c.
\newblock {\em ORTEC}.

\bibitem{UTROBICIC201521}
A.~Utrobicic, M.~Kovacic, F.~Erhardt, N.~Poljak, and M.~Planinic.
\newblock A floating multi-channel picoammeter for micropattern gaseous
  detector current monitoring.
\newblock {\em Nucl. Instrum. Meth. A}, 801:21--26, 2015.

\bibitem{MORMANN2004315}
D~Mörmann, A~Breskin, R~Chechik, and D~Bloch.
\newblock Evaluation and reduction of ion back-flow in multi-gem detectors.
\newblock {\em Nucl. Instrum. Meth. A}, 516(2):315--326, 2004.

\bibitem{KETZER2013237}
Bernhard Ketzer.
\newblock A time projection chamber for high-rate experiments: Towards an
  upgrade of the alice tpc.
\newblock {\em Nuclear Instruments and Methods in Physics Research Section A:
  Accelerators, Spectrometers, Detectors and Associated Equipment},
  732:237--240, 2013.
\newblock Vienna Conference on Instrumentation 2013.

\bibitem{BARBOSA2019162356}
F.{Barbosa et. al.}
\newblock A new transition radiation detector based on gem technology.
\newblock {\em Nuclear Instruments and Methods in Physics Research Section A:
  Accelerators, Spectrometers, Detectors and Associated Equipment}, 942:162356,
  2019.

\bibitem{SAULI2006269}
F.~Sauli, L.~Ropelewski, and P.~Everaerts.
\newblock Ion feedback suppression in time projection chambers.
\newblock {\em Nucl. Instrum. Meth. A}, 560(2):269--277, 2006.

\bibitem{Luz_2018}
H.~Natal da~Luz, P.~Bhattacharya, L.A.S. Filho, and L.E.F.M. Fran{\c{c}}a.
\newblock Ion backflow studies with a triple-{GEM} stack with increasing hole
  pitch.
\newblock {\em Journal of Instrumentation}, 13(07):P07025--P07025, jul 2018.

\bibitem{Ball_2014}
M~Ball, K~Eckstein, and T~Gunji.
\newblock Ion backflow studies for the {ALICE} {TPC} upgrade with {GEMs}.
\newblock {\em Journal of Instrumentation}, 9(04):C04025--C04025, apr 2014.

\bibitem{KILLENBERG2004251}
M.~{Killenberg et al.}
\newblock Charge transfer and charge broadening of gem structures in high
  magnetic fields.
\newblock {\em Nuclear Instruments and Methods in Physics Research Section A:
  Accelerators, Spectrometers, Detectors and Associated Equipment},
  530(3):251--257, 2004.

\bibitem{TARAFDAR2022167460}
Sourav Tarafdar, Senta~V. Greene, Julia Velkovska, Brandon Blankenship, and
  Michael~Z. Reynolds.
\newblock Reduction of ion backflow using a quadruple gem detector with various
  gas mixtures.
\newblock {\em Nuclear Instruments and Methods in Physics Research Section A:
  Accelerators, Spectrometers, Detectors and Associated Equipment}, page
  167460, 2022.

\end{thebibliography}


\begin{thebibliography}{1}
\bibitem{GEM_sauli}
Fabio Sauli
\newblock GEM: A new concept for electron amplification in gas detectors
\newblock In {\em Nuclear Instruments and Methods in Physics Research Section A: Accelerators, Spectrometers, Detectors and Associated Equipment, volume 386}, pages 531-534. 1994.

\bibitem{AIDALA2003200}
C.~Aidala, B.~Azmoun, Z.~Fraenkel, T.~Hemmick, B.~Khachaturov, A.~Kozlov,

A.~Milov, I.~Ravinovich, I.~Tserruya, S.~Stoll, C.~Woody, and S.~Zhou.
\newblock A hadron blind detector for phenix.
\newblock {\em Nuclear Instruments and Methods in Physics Research Section A:
  Accelerators, Spectrometers, Detectors and Associated Equipment}, 502(1):200
  -- 204, 2003.
\newblock Experimental Techniques of Cherenkov Light Imaging. Proceedings of
  the Fourth International Workshop on Ring Imaging Cherenkov Detectors.

\bibitem{GARABATOS2004197}
C.~Garabatos.
\newblock The alice tpc.
\newblock {\em Nuclear Instruments and Methods in Physics Research Section A:
  Accelerators, Spectrometers, Detectors and Associated Equipment}, 535(1):197
  -- 200, 2004.
\newblock Proceedings of the 10th International Vienna Conference on
  Instrumentation.
\bibitem{Tarafdar:2019oB}
Sourav Tarafdar.
\newblock {sPHENIX TPC simulation studies}.
\newblock {\em PoS}, MPGD2017:067, 2019.

\bibitem{GEM_sauli2}
Fabio Sauli.
\newblock The gas electron multiplier (GEM): Operating principles and
  applications.
\newblock {\em Nuclear Instruments and Methods in Physics Research Section A:
  Accelerators, Spectrometers, Detectors and Associated Equipment}, 805:2 --
  24, 2016.

\end{thebibliography}

\end{document}